\documentclass[aps, prb, superscriptaddress,preprintnumbers,amssymb, preprint]{revtex4-2}

\usepackage{graphicx}
\usepackage{dcolumn}
\usepackage{bm}
\usepackage{amsmath, amssymb}
\usepackage{newtxtext, newtxmath}
\usepackage{mathrsfs}
\usepackage{comment}

\DeclareMathOperator{\e}{e}
\newcommand{\ud}{\mathrm{d}}
\def\degree{\kern-.2em\r{}\kern-.3em}

\begin{document}

\preprint{Ver. 11p1}

\title{In-plane electronic anisotropy revealed by interlayer resistivity measurements on the iron-based superconductor parent compound CaFeAsF}


\author{Taichi Terashima}
\email{TERASHIMA.Taichi@nims.go.jp}
\affiliation{International Center for Materials Nanoarchitectonics, National Institute for Materials Science, Tsukuba, Ibaraki 305-0003, Japan}
\author{Hishiro T. Hirose}
\affiliation{Research Center for Functional Materials, National Institute for Materials Science, Tsukuba, Ibaraki 305-0003, Japan}
\author{Yoshitaka Matsushita}
\affiliation{Research Network and Facility Services Division, National Institute for Materials Science, Tsukuba, Ibaraki 305-0047, Japan}
\author{Shinya Uji}
\affiliation{International Center for Materials Nanoarchitectonics, National Institute for Materials Science, Tsukuba, Ibaraki 305-0003, Japan}
\author{Hiroaki Ikeda}
\email{hikeda.uji@gmail.com}
\affiliation{Department of Physics, Ritsumeikan University, Kusatsu, Shiga 525-8577, Japan}
\author{Yuki Fuseya}
\affiliation{Department of Engineering Science, University of Electro-Communications, Chofu, Tokyo 182-8585, Japan}
\author{Teng Wang}
\author{Gang Mu}
\email{mugang@mail.sim.ac.cn}
\affiliation{State Key Laboratory of Functional Materials for Informatics, Shanghai Institute of Microsystem and Information Technology, Chinese Academy of Sciences, Shanghai 200050, China}
\affiliation{CAS Center for Excellence in Superconducting Electronics (CENSE), Shanghai 200050, China}

\date{\today}

\begin{abstract}
Both cuprates and iron-based superconductors demonstrate nematicity, defined as the spontaneous breaking of rotational symmetry in electron systems.
The nematic state can play a role in the high-transition-temperature superconductivity of these compounds.
However, the microscopic mechanism responsible for the transport anisotropy in iron-based compounds remains debatable.
Here, we investigate the electronic anisotropy of CaFeAsF by measuring its interlayer resistivity under magnetic fields with varying field directions.
Counterintuitively, the interlayer resistivity was larger in the longitudinal configuration ($B \parallel I \parallel c$) than in the transverse one ($B \perp I \parallel c$).
The interlayer resistivity exhibited a so-called coherence peak under in-plane fields and was highly anisotropic with respect to the in-plane field direction. At $T$ = 4 K and $B$ = 14 T, the magnetoresistance $\Delta\rho/\rho_0$ was seven times larger in the $B \parallel b_o$ than in the $B \parallel a_o$ configuration.
Our theoretical calculations of the conductivity based on the first-principles electronic band structure qualitatively reproduced the above observations but underestimated the magnitudes of the observed features.
The proposed methodology can be a powerful tool for probing the nematic electronic state in various materials.
\end{abstract}


\maketitle
\newpage



\section{introduction}
The parent compounds of iron-based superconductors typically exhibit a tetragonal-to-orthorhombic structural transition at temperature $T_s$ [Fig. 1(a)], which is equal to or slightly higher than the antiferromagnetic transition temperature $T_N$.
When the two transitions are suppressed by chemical substitution or pressure application, these compounds exhibit superconductivity \cite{Kamihara08JACS, Rotter08PRL, Sasmal08PRL, Johnston10AdvPhys}.
How the two transitions are related to superconductivity is currently being debated.

This structural transition has been proposed as a nematic transition of electronic origin because the in-plane resistivity is significantly anisotropic below $T_s$ despite a tiny orthorhombic distortion \cite{Chu10Science}.
To probe the nematic fluctuations above the transition temperature, Chu \textit{et al}.\cite{Chu12Science} determined the elastoresistance of Ba(Fe$_{1-x}$Co$_x$)$_2$As$_2$, which defines the resistance change under strain.
Evidence of the nematic electronic state was corroborated by the relevant band-energy shift in angle-resolved photoemission spectroscopy \cite{Yi11PNAS}, a unidirectional structure around impurities (dubbed ``nematogens'') in scanning tunneling microscopy images \cite{Chuang10Science}, and nematic fluctuations in Raman scattering data \cite{Gallais13PRL}.

Although nematicity has been well established in iron-based compounds, the microscopic mechanism responsible for the transport anisotropy remains contentious.
In particular, whether the anisotropy arises from Fermi surface anisotropy or scattering phenomena is unclear.
Optical studies favor the former origin \cite{Mirri15PRL}, whereas annealing and doping effects suggest the latter \cite{Nakajima11PNAS, Blomberg12NatCommun}.
A real-space picture based on nematogens has also been proposed \cite{Nakajima12PRL, Gastiasoro14PRL}.
To gain further insight into this fundamental issue, we apply here a new methodology to a compound whose nematicity has not been previously studied.

Most published nematicity studies have been performed on 122-type iron arsenides such as BaFe$_2$As$_2$ because they form large, high-quality single crystals.
In addition, angle-resolved photoemission spectroscopy of 1111-type arsenides such as LaFeAsO suffers from contamination by surface electronic structures \cite{Liu10PRB}.
The present study focuses on CaFeAsF, a variant of the 1111-type arsenides with the same ZrCuSiAs-type structure as LaFeAsO but with a CaF layer replacing the LaO layer of LaFeAsO \cite{Matsuishi08JACS}.
High-quality single crystals exhibiting quantum oscillations can be grown using the flux method \cite{Ma15SST, Terashima18PRX}.
CaFeAsF exhibits a nonmetallic temperature dependence of electrical conduction, i.e., d$\rho$/d$T < 0$, from room temperature down to $T_s$. 
In contrast, the 122 compounds show metallic conduction as the temperature decreases from room temperature.
The Fermi surface of CaFeAsF in the antiferromagnetic state below $T_N$ is quasi-two dimensional (Q2D), being composed of a tiny hole cylinder at the zone center surrounded by a pair of symmetrically arranged tiny Dirac electron cylinders [Fig. 1(a)].
This structure contrasts with the three-dimensional Fermi surface of the 122 compounds in the antiferromagnetic state, which is composed of closed pockets\cite{Terashima11PRL}.
These differences between CaFeAsF and the 122 compounds highlight the importance of studying nematicity in CaFeAsF.

Our main methodology is based on interlayer resistivity measurements under an applied magnetic field.
Since the discovery of angle-dependent magnetoresistance oscillations in organic conductors \cite{Kartsovnik88JETPLett, Kajita89SSC}, interlayer resistivity measurements have become a powerful tool of fermiology for Q2D electron systems, as exemplified by their application to cuprates \cite{Hussey03Nature, Grissonnanche21Nature}.
Contrary to usual expectations, we found that the magnetoresistance of the interlayer resistivity in CaFeAsF is larger in the longitudinal configuration ($B \parallel I \parallel c$) where the Lorentz force acting on electrons is expected to be minimal than in the transverse one ($B \perp I \parallel c$).
We observed a coherence peak under in-plane magnetic fields, which strongly depends on the in-plane field direction.
To calculate the conductivity, we applied Chambers' expression to the electronic band structure determined using first-principles calculations.
The calculated results qualitatively reproduce the above experimental observations.
However, the magnitudes of the calculated features were weaker than the experimentally observed magnitudes.
We discuss possible origins of this quantitative discrepancy.

The paper is organized as follows:
We begin with measurements of in-plane resistivity and elastoresistance in Sec. II A to demonstrate the nematicity in CaFeAsF.
In Sec. II B, we present results of interlayer resistivity.
We perform theoretical conductivity calculations in Sec. II C.
We discuss the results in Sec. III.
Details of experimental procedures and theoretical calculations are described in Appendices A and B, respectively.

\section{Results}
\subsection{In-plane resistivity and elastoresistance}
First, we established nematicity in CaFeAsF from in-plane resistivity and elastoresistance measurements.
Figure 1(b) shows the in-plane resistivity (green curve) measured on a free-standing sample and the elastoresistance (black curve) of the same sample.
The sample was bar-shaped with its longest dimension along the tetragonal [110] direction ([110]$_t$ where the subscript $t$ indicates a tetragonal cell). 
We applied the electrical current and strain along this direction (see Appendix A for details of the experiments).
The strain was applied using a piezostack.
From $\ud^2\rho/dT^2$ (light-blue curve), we obtained $T_s=116.8$ K and $T_N = 106.3$ K for this sample.
The resistivity gradually increased as the temperature decreased from room temperature to $T_s$, but it decreased sharply below $T_s$.
The negative elastoresistance shows that the resistivity decreased with the elongation of the sample, as also observed in BaFe$_2$As$_2$ \cite{Chu12Science} and LaFeAsO \cite{Hong20PRL}.
The magnitude of the elastoresistance increased as the temperature decreased to $T_s$.
From the Curie--Weiss fit to the data between 200~K and $T_s$ (red dotted curve), we determined the Weiss temperature to be 103.9 K, close to $T_N$.
Although twinning prevents a straightforward interpretation of the elastoresistance data below $T_s$, the elastoresistance exhibited a kink at $T_N$ which was
absent in the data reported for Ba(Fe$_{1-x}$Co$_x$)As$_2$ and La(Fe$_{1-x}$Co$_x$)AsO \cite{Chu12Science, Hong20PRL}.

Figure 1(c) compares the in-plane resistivities of another sample before and after fixing it to a polyetheretherketone (PEEK) substrate.
This sample was also bar-shaped, with its length oriented along the [110]$_t$ direction.
As the free-standing sample is expected to be heavily twinned when cooled below $T_s$, the resistivity measured before fixing it to the PEEK substrate (green curve) corresponds to the average of the resistivities along the $a_o$ and $b_o$ axes of the orthorhombic cell (denoted by the subscript $o$).
When both ends of the sample are fixed to the PEEK substrate [see Fig. 2(a)], the substrate shrinks more than the sample when cooled, so the [110]$_t$ direction becomes the shorter $b_o$ direction of the orthorhombic cell below $T_s$ through most of the sample volume. 
Therefore, the resistivity measured after fixing [pink curve, Fig. 1(c)] corresponds approximately to the resistivity of the $b_o$ axis.
Figure 1(d) shows the in-plane resistivity of a third sample before (green) and after (amber) it is fixed to a quartz substrate.
In this case, the cooled substrate does not shrink; thus, the resistivity of the longer $a_o$ axis is approximately measured below $T_s$.
Regardless of substrate, the resistivity peak at $T_s$ is broadened considerably after fixing because the stress along the [110]$_t$ axis enforces a finite nematic order parameter above $T_s$, analogously to a magnetic field applied to a ferromagnet.
Figures 1(c) and (d) also show the normalized differences $\Delta \rho/\rho = (\rho_{\mathrm{fixed}} - \rho_{\mathrm{free}})/ \rho_{\mathrm{free}}$.
The sign and rapid magnitude increase of the normalized difference $\Delta \rho/\rho$ below $\sim$200 K are consistent with elastoresistance.
The magnitude of the normalized difference increased further below $T_s$, showing a kink at $T_N$.
Assuming that both samples were completely detwinned, we estimate $\rho_{b_o}/\rho_{a_o}$ to be 2.2 at $T$ = 4.2 K, much larger than the value reported for BaFe$_2$As$_2$ \cite{Chu12Science}.

There are some differences between the free-standing resistivity curves (green) in Figs. 1(b)--(d).
Although the exact origins are unclear, possible origins include the following:
As the elastoresistance is large, the resistivity is affected by internal strain, which varies from sample to sample.
In addition, free-standing samples are twinned below $T_s$, and hence the resistivity is a mixture of the $a_o$-axis and $b_o$-axis resistivities and depends on distribution of domains.

\subsection{Interlayer resistivity}
We now investigate the interlayer resistivity.
Figure 2(a) and the inset of Fig. 2(b) show an image and schematic of the sample, respectively, on which the interlayer resistivities $\rho_c$ were measured before and after fixing the sample to the PEEK substrate [Fig. 2(b)].
From $\ud^2\rho_c/dT^2$ of the free-standing sample (green curve), we obtained $T_s=117.6$ K and $T_N = 106.8$ K as indicated by broken vertical lines.
After fixing this sample to the substrate, its interlayer resistivity decreased (black curve), and the peak at $T_s$ broadened considerably, as observed for the in-plane resistivity, but the transition at $T_N$ remained sharp, showing only a slight upward temperature shift to $T_N^{\prime}$ = 110.8 K.

Figure 2(c) compares the zero-field resistivity (black line) and the interlayer resistivities measured under an applied field of $B$ = 14 T along the $a_o$, $b_o$, and $c$ axes (squares, circles, and crosses, respectively) (the subscript is omitted from the $c$ axis because this axis is identical in the tetragonal and orthorhombic phases).
The magnetoresistance was negligible at 140 K but began emerging approximately below $T_s$ of the free-standing sample.
As the temperature was lowered, considerable magnetoresistance developed under $B \parallel c$ and $B \parallel b_o$, but magnetoresistance was much smaller under $B \parallel a_o$.
The pink line in Fig. 2(c) plots the resistivity difference between $B \parallel b_o$ and $B \parallel a_o$ (a ten-times enlarged curve is also shown between 90 and 140 K).
The finding of maximal magnetoresistance along $B \parallel c$ (i.e., in the longitudinal configuration) is counterintuitive because electron motion along the magnetic-field direction is unaffected by the Lorentz force; accordingly, the magnetoresistance for $I \parallel c$ should be minimized under the $B \parallel c$ condition.
Similar counterintuitive observations of interlayer resistivity have been reported in other Q2D electron systems such as organic conductors \cite{Clarke98Science, Kartsovnik06PRL}, high-$T_c$ cuprates \cite{Hussey98PRB}, and SrMnBi$_2$ \cite{Jo14PRL}. 
Such observations are often interpreted as incoherent interlayer transport.
Also remarkable is the difference between the two in-plane field orientations, namely, $B \parallel b_o$ and $B \parallel a_o$.
This anisotropy exists even above $T_N^{\prime}$:
Figures 2(e) and (f) show the magnetic-field dependence of the interlayer resistivity at $T$ = 112 and 115 K ($> T_N^{\prime}$) for three field directions.
Because the magnetoresistance is small, the quantization error is visible.
Although the zero-field resistivity differs between the magnetoresistance curves, the differences are only 0.06\%, which may be ascribed to measurement inaccuracy or a temperature difference of 0.04 K.
While the magnetoresistance is apparent for $B \parallel b_o$ (red line), it is negligible for $B \parallel a_o$ (blue line) at these temperatures.
At $T$ = 140 K, sufficiently above $T_s$ of the free-standing sample, the magnetoresistance is negligibly small (-0.08\% at $B$ = 14 T) for all the field directions [Fig. 2(g)].

Figure 2(d) plots the interlayer resistivity at $T$ = 4 K as a function of $B$ applied parallel to the $a_o$, $b_o$, and $c$ axes.
At any field strength, applying $B \parallel c$ maximized the magnetoresistance.
The resistivity curve under $B \parallel c$ was weakly concave downward except under low fields ($B \lesssim 0.6$ T) and except when the resistivity was affected by the Shbunikov--de Haas (SdH) effect, which caused wiggling behavior above $\sim$8 T.
The concave-downward curve suggests eventual saturation, although saturation can never be observed because the sample undergoes a metal–insulator transition at 30 T under $B \parallel c$ \cite{Ma18scichina, Terashima22npjQM2}.
In contrast, the resistivities under $B \parallel a_o$ and $b_o$ increased with no sign of saturation.
The resistivity is larger for $B \parallel b_o$ than for $B \parallel a_o$ at any field magnitude.

We now present our central experimental results.
Figure 3(a) shows the magnetic-field-direction dependence of the interlayer resistivity at $T$ = 4 K and $B$ = 14 T.
In these measurements, the polar angle $\theta$ was varied while the azimuthal angle $\phi$ remained constant [see Fig. 3(b)].
Tilting the field from the $c$ axis ($\theta$ = 0) decreased the interlayer resistivity, consistent with Fig. 2(d).
Angle-dependent magnetoresistance oscillations were not observed because the Fermi wave vector $k_F$ is very small \cite{Terashima18PRX}.
As $\theta$ approached $\pm$90$^{\circ}$, the $\rho_c(\theta)$ curves exhibited a peak that was hardly visible near $\phi = 0$ and 180$^{\circ}$ ($B \parallel a_o$) but was pronounced near $\phi = 90^{\circ}$ ($B \parallel b_o$).
Figure 3(c) shows an azimuthal equidistant projection of the same data (the projection method is the same as that used for generating the world map on the emblem of the United Nations).
The map evinces a clear two-fold symmetry.
Panels (d) and (e) of Fig. 3 show the field and temperature dependences of the $\rho_c(\theta)$ curves at $\phi$ = 0 and 90$^{\circ}$, respectively.
The curves are almost sinusoidal, meaning that the magnetoresistance (except for the peak at $\theta$ = $\pm$90$^{\circ}$) was essentially determined by the $c$-axis component of the magnetic field.
The obvious deviation from the sinusoidal form seen at $T \leqslant 4$ K and $B$ = 14 T is attributable to the SdH effect.
The magnetoresistance was maximized at $B \parallel c$ ($\theta = 0$) at all temperatures up to $T$ = 115 K [see also Fig. 2(f)].
The peak at $\theta$ = $\pm$90$^{\circ}$ appeared up to $T$ = 10 and 70 K at $\phi$ = 0 and 90$^{\circ}$, respectively.
Whereas the peak height decreased with decreasing $B$ or increasing $T$, the peak width showed no appreciable change.
Figure 3(f) plots the interlayer resistivities as functions of $\phi$ at $T$ = 4 K under various constant in-plane fields ($\theta$ = 90$^{\circ}$).
The solid curves were fitted to
\begin{equation}
(\sigma_{1}/(1+r\sin^{2}\phi)+\sigma_{2})^{-1}.
\end{equation}
\cite{Zhu11NatPhys}.
The perfect fits affirm that the sample was almost completely detwinned.
Imperfect detwinning would manifest as flattened peaks or a local minimum at $\phi = \pm90^{\circ}$ caused by admixture of the $B \parallel a_o$ component.
Figure 3(g) is a polar plot of the $B$ = 14 T data in Fig. 3(f).
The plot confirms two-fold symmetry with a pronounced anisotropy.

\subsection{Theoretical conductivity}

We now present our theoretical calculations of magnetoresistance in CaFeAsF.
To this end, we evaluated Chambers' expression for the conductivity tensor \cite{Chambers52ProcPhysSocA}, a solution of the Boltzmann equation in the relaxation-time approximation, and adopted the relativistic electronic band structure calculated in \cite{Terashima18PRX} (see Appendix B for details of the calculations).
The electron and hole conductivities, $\sigma^{e}$ and $\sigma^{h}$, respectively, were computed separately.
The $x$, $y$, and $z$ axes were taken along the $a_o$, $b_o$, and $c$ axes, respectively.
In the calculated results, the approximation $\rho_{zz}=1/\sigma_{zz}$ was found to hold accurately. 
We therefore defined the electron (hole) resistivities along the $z$ axis as $\rho_{zz}^{e(h)}=1/\sigma_{zz}^{e(h)}$.

We first note that under the constant-$\tau$ approximation, $\rho_{yy}/\rho_{xx}$ was 2.4 at zero-field, consistent with the experimental anisotropy of 2.2.
This result indicates that the anisotropy of the Fermi surface and velocity is basic to the explanation of the observed resistivity anisotropy.

The calculated magnetoresistance $\Delta\rho/\rho_{0}$ was one order of magnitude larger for the electrons than for the holes.
Accordingly, we consider only the electron resistivity in the following analysis.
The qualitative features of magnetoresistance remain identical when the hole resistivity is included.
Figure 4(a) shows the magnetic-field dependences of $\rho_{zz}^{e}$ in the three field directions.
Under weak fields ($B < 14.2$~T), the largest resistivity was $\rho_{zz}^{e}$ at $B \parallel z$ [hereafter denoted as $\rho_{zz}^{e}(B \parallel z)$], which was concave downward except at very low fields ($B \leqslant 0.3$~T).
In addition, $\rho_{zz}^{e}(B \parallel y) > \rho_{zz}^{e}(B \parallel x)$.
These results are consistent with the experimental observations.
Above $B \sim 5$~T, $\rho_{zz}^{e}(B \parallel z)$ tended to saturate but $\rho_{zz}^{e}(B \parallel y)$ and $\rho_{zz}^{e}(B \parallel x)$ did not show such a tendency.
This result explains why $\rho_{zz}^{e}(B \parallel y)$ exceeded $\rho_{zz}^{e}(B \parallel z)$ at $B > 14.2$~T and also suggests that $\rho_{zz}^{e}(B \parallel x)$ will exceed $\rho_{zz}^{e}(B \parallel z)$ at yet higher fields.
Thus, the intuitive conjecture of that the transverse magnetoresistance [i.e., $\rho_{zz}^{e}(B \parallel x)$ and $\rho_{zz}^{e}(B \parallel y)$] will be larger than the longitudinal magnetoresistance [$\rho_{zz}^{e}(B \parallel z)$] is correct in the high-field limit.
The experimental observations of the larger longitudinal magnetoresistance can be explained without invoking incoherent interlayer transport.
It is because the high-field limit was not reached in the present experimental conditions.
The crossing of $\rho_{zz}^{e}(B \parallel c)$ and $\rho_{zz}^{e}(B \parallel b_o$ or $a_o$) was not experimentally observed up to 14 T [Fig. 2(d)].

Figure 4(b) plots $\rho_{zz}^{e}$ at $B = 13$~T as a function of $\theta$ and $\phi$ near $\theta$ = 90 $^{\circ}$ (we chose 13 T, avoiding the crossing of the $\rho_{zz}^{e}(B \parallel z)$ and $\rho_{zz}^{e}(B \parallel y)$ curves).
A clear resistivity peak developed at $\theta$ = 90$^{\circ}$ as the field direction approached $\phi$ = 90$^{\circ}$ ($B \parallel b_o$).
Figure 4(c) plots the resistivity at $\theta$ = 90$^{\circ}$ as a function of $\phi$ (circles).
The data were well-fitted by Eq. 1 (solid line).
These theoretical results are in excellent qualitative agreement with the experimental data.
However, we note that the magnitude of the magnetoresistance is greatly underestimated in the present calculations:
the calculated values of $\Delta \rho / \rho_0$ are 4 and 10\% at $B$ = 13 T for $B \parallel a_o$ and $b_o$, respectively, compared to the experimental values of 25 and 181\% at $B$ = 14 T.

The interlayer conductivity in the antiferromagnetic phase of iron-pnictides under magnetic fields has been studied theoretically \cite{Morinari09JPSJ}.
The $\phi$ dependence of the magnetoresistance in \cite{Morinari09JPSJ} is the opposite of our study because they adopted a different electronic-structure model.

\section{Discussion}
The peak in the $\rho_c(\theta)$ curve at $\theta$ = 90$^{\circ}$ is reminiscent of the interlayer coherence peak, which was initially found in organic conductors \cite{Kartsovnik88JETPLett} and later observed in other Q2D electron systems such as Sr$_2$RuO$_4$ \cite{Ohmichi99PRB}, SrMnBi$_2$ \cite{Jo14PRL}, and KFe$_2$As$_2$ \cite{Kimata10PRL}.
This peak has been ascribed to small closed orbits \cite{Hanasaki98PRB} or self-crossing orbits \cite{Peschansky99PRB} on the sides of Fermi cylinders when the field is parallel to the conducting layers.
Its width is determined by the magnitude of the interlayer dispersion relative to the Fermi energy and is independent of field strength and temperature.
Consistent with these reports, the width of the experimentally observed peak showed no appreciable dependence on the magnetic field or temperature [Figs. 3(d) and (e), respectively].
The peak width, defined as the distance from the peak to the resistivity minimum on either side, was approximately 10$^{\circ}$ under $B \parallel b_o$, much larger than the order-1$^{\circ}$ peak widths typically found in organic conductors \cite{Kartsovnik88JETPLett, Hanasaki98PRB} but consistent with our theoretical electron resistivity [which yielded a peak width of 9$^{\circ}$ at $\phi$ = 90$^{\circ}$; see Fig. 4(b)].
The large peak width reflects a relatively large interlayer transfer in the present case.

We now consider the in-plane field-angle dependence of the resistivity at $\theta$ = 90 $^{\circ}$ [Figs. 3(f) and 4(c)].
Within the (semi-)classical approximation, the conductivity tensor is given by $\hat{\sigma} = ne(\hat{\mu}^{-1} \pm \hat{B})^{-1}$, where $\hat{\mu}$ is the mobility tensor and $+$ ($-$) denotes holes (electrons) \cite{Mackey69PRB, Mitani20JPCM}.
The magnetic tensor is given by
\begin{equation}
\hat{B} =
\left(\begin{array}{ccc}
0 & -B_z & B_y \\
B_z & 0 & -B_x \\
-B_y & B_x & 0 \\
\end{array} \right).
\end{equation}
Taking the $a_o$, $b_o$, and $c$ axes as the $x$, $y$, and $z$ axes, respectively, and assuming in-plane fields $(B_x, B_y, 0) = (B\cos\phi, B\sin\phi, 0)$, we find the interlayer conductivity to be
\begin{equation}
\sigma_{zz} = ne\mu_z \frac{1}{1 + \mu_y\mu_zB_x^2 + \mu_z\mu_xB_y^2}.
\end{equation}
This expression can be rewritten as
\begin{equation}
\sigma_{zz} = \frac{\sigma_1}{1 + r\sin^2\phi},
\end{equation}
where
\begin{eqnarray}
\sigma_1 &=& \frac{ne\mu_z}{1+\mu_y\mu_zB^2},\ \mathrm{and}\\
r &=& \frac{\mu_y\mu_zB^2}{1+\mu_y\mu_zB^2}\left(\frac{\mu_x}{\mu_y}-1\right).
\end{eqnarray}
The above fitting function Eq. 1 was obtained by adding another conduction channel $\sigma_2$ independent of the field direction $\phi$ and assuming $\rho_{zz} = \sigma_{zz}^{-1}$.
From Fig. 3(f), we have $r > 0$ and hence $\mu_x > \mu_y$, consistent with the zero-field in-plane resistivity data (Fig. 1).

We now relate the $\phi$ dependence to the Fermi-surface anisotropy.
According to theoretical studies of magnetoresistance in organic conductors, the interlayer conductivity of Q2D electron systems under an in-plane magnetic field is proportional to \cite{Lebed97PRB, Peschansky97LTP}
\begin{equation}
\oint\frac{dl}{|\bm{\varv}_n|[1+(C\tau B |\bm{\varv}_n|\sin\alpha)^2]},
\end{equation}
where $dl$ is a line element along a cross-section of the Fermi surface normal to the $c$ axis, $\bm{\varv}_n$ defines the in-plane component of the Fermi velocity, $C$ is a prefactor depending on the interlayer distance, and $\alpha$ is the angle between $\bm{\varv}_n$ and the field.
Note that $B |\bm{\varv}_n|\sin\alpha$ is proportional to the Lorentz force along the $k_z$ direction (the first term in the denominator, $|\bm{\varv}_n|^{-1}$, is a density-of-states factor).
If the Lorentz force is large, the electrons rapidly traverse the $\mathbf{k}$ space in the $k_z$ direction, causing rapid oscillations of the interlayer velocity around zero \cite{Kartsovnik04CR}.
This behavior diminishes the electrons' contribution to the interlayer conductivity.
According to the above formula, interlayer conduction is mainly contributed by electrons located at $\mathbf{k}$ points where $\bm{\varv}_n$ is nearly parallel to the field.
Now imagine that the in-plane shape of the Fermi surface is an ellipse elongated along the $k_y$ direction.
Under an in-plane magnetic field, the interlayer conductivity is larger at $B \parallel k_x$ than at $B \parallel k_y$ because more electrons are located where $\bm{\varv}_n$ is nearly parallel to the applied field.
According to \cite{Lebed97PRB}, magnetoresistance is approximately linear in a magnetic field and the ratio of the slopes (i.e., d$\rho_{zz}$/d$B$) for $B \parallel k_x$ and $k_y$ equals the ratio of the Fermi wave vectors along $k_x$ and $k_y$.

The cross-sections of the calculated electron and hole Fermi cylinders at $k_z$ = 0 are shown in Fig. 5.
Both cross-sections are elongated along the $k_y$ axis, consistent with the larger magnetoresistance observed at $B \parallel b_o$.
However, the small aspect ratios of the electron and hole pockets (1.4 and 1.7, respectively) are incompatible with the observed pronounced difference between $B \parallel a_o$ and $b_o$ within the framework of the above theory. 
The ratio of the experimental magnetoresistance slopes d$\rho_c$/d$B$ for $B \parallel a_o$ and $b_o$ is as large as 5.6 at $B$ = 14 T [Fig. 2(d)].
Interestingly, the theoretical calculations yield a ratio of 2.1 at $B$ = 13 T [Fig. 4(a)], which is in a fair agreement with the small aspect ratios.

The present theoretical conductivity calculations qualitatively explain the field-direction anisotropy and coherence peak observed in the experiment, whereas they considerably underestimate the magnitudes of the magnetoresistance and anisotropy.
Therefore, factors beyond the present constant-$\tau$ calculations must be considered.
First, the above discrepancy between the experimental slope ratio and aspect ratio may indicate that the Fermi pockets are more elongated along the $b_o$ axis than predicted by the band-structure calculation. 
However, this conjecture may be incompatible with the fact that the present conductivity calculations reasonably reproduced the zero-field anisotropy $\rho_{b_o}/\rho_{a_o}$.
Second, distribution of orbital contents on the Fermi surface should be taken into account.
The in-plane-field anisotropy of the magnetoresistance exists above $T_N^{\prime}$, where only the nematic order parameter is finite [Figs. 2(e) and (f)].
Because the orthorhombic distortion is slight, the Brillouin zone in this state, i.e., $T_N < T < T_s$, is nearly identical to that in the tetragonal phase, and band-folding is not expected.
However, the nematic order parameter lifts the degeneracy of the $yz$ and $zx$ orbitals \cite{Fernandes14NatPhys, Yamakawa16PRX}.
The nematic order could also involve the $xy$ orbital.
An optical study of CaFeAsF has found a sharp decrease in the Drude weight at $T_s$ \cite{Xu18PRB}, confirming a radical change in the electronic structure owing to rearranging the $yz$, $zx$, and $xy$ orbital weights near the Fermi level.
Thus, the emergence of the in-plane-field anisotropy above $T_N^{\prime}$ can be ascribed to the rearranged orbital contents on the Fermi surface.
The orbital differentiation is more dramatic in the antiferromagnetic sate ($T < T_N$), which is most likely related to an enhancement of the in-plane-filed anisotropy across $T_N^{\prime}$ [see the ten-times enlarged difference curve (pink) in Fig. 2(c)].
Our theoretical calculations were performed in this antiferromagnetic state.
In the antiferromagnetic state below $T_N$, the electron pockets appear at Dirac points, which are the crossing points of the $t_{2g}$ (i.e., $yz$, $zx$, and $xy$) and the $e_g$ (i.e., $x^2-y^2$ and $3z^2-r^2$) band (without spin-orbit coupling).
Accordingly, the right side of the electron pocket in Fig. 5 is dominated by the $zx$ orbital with a moderate admixture of $xy$, whereas the left side is dominated by $x^2-y^2$ and $3z^2-r^2$ orbitals.
As indicated on the velocity map, the $\varv_x$ component of the Fermi velocity is large on the right side of the electron pocket and thus suppresses the interlayer conductivity under $B \parallel b_o$ (see Eq. 7).
Furthermore, because the orbital contents differ along the cross-sections, scattering is intraorbital or interorbital depending on from where to where electrons are scattered.
For example, intrapocket scattering within an electron pocket along $k_x$ and $k_y$ is primarily interorbital and intraorbital, respectively.
Scattering between the hole pocket and the $e_g$-side of the electron pockets is interorbital because the $xy$ orbital dominates the hole pocket.
This degrades the appropriateness of the constant-$\tau$ approximation, and the anisotropy between $B \parallel a_o$ and $B \parallel b_o$ could be enhanced.

We may also need to consider the antiferromagnetic order and its response to applied magnetic fields.
Below $T_N$, the Fe spins in CaFeAsF are aligned in the $a_o$ direction and coupled antiferromagnetically along the $a_o$ and $c$ axes but ferromagnetically along the $b_o$ axis \cite{Xiao09PRB_CaFeAsF}.
One might argue that magnetic fields cant the spins and hence alter the electronic structure.
Since the magnetic susceptibility along the $b_o$ axis corresponds to a perpendicular susceptibility of an antiferromagnet, the $b_o$-axis susceptibility is expected to be larger than the  $a_o$-axis one.
This might suggest that the electronic structure is more susceptible to the $b_o$-axis field and hence might explain the larger magnetoresistance for $B \parallel b_o$ than $a_o$ [Fig. 2 (d)].
However, we note that previous quantum oscillation measurements found no evidence of magnetic-field induced changes in the electronic structure \cite{Terashima18PRX, Terashima22npjQM, Terashima22npjQM2}:
the oscillations observed in a field range up to $\sim$20 T and a field-angle $|\theta| \lesssim 70^{\circ}$ conformed to the standard Lifshitz-Kosevich formula, which assumes no change in the electronic structure except for trivial Zeeman energy of electron spins \cite{Shoenberg84}.
One might also argue that magnetic anisotropy leads to anisotropic magnetoresistance via anisotropic magnetic fluctuations.
However, we note that magnetic fluctuations are suppressed as the temperature is lowered:
accordingly, magnetoresistance anisotropy would also diminishes.
However, the experimental magnetoresistance anisotropy was enhanced as the temperature was lowered [Fig. 2(c)].

In summary, we identified three notable features in our interlayer resistivity measurements on CaFeAsF:
(1) The magnetoresistance was maximized in the longitudinal configuration ($I \parallel B \parallel c$).
(2) A coherence peak appeared at $\theta$ = 90$^{\circ}$.
(3) The interlayer resistivity under a constant in-plane field strongly depended on $\phi$.
The magnetoresistance slope d$\rho_c$/d$B$ is more than five times larger for $B \parallel b_o$ than for $B \parallel a_o$ at $T$ = 4 K and $B$ = 14 T.

Our theoretical calculations within the constant-$\tau$ approximation qualitatively reproduced these features, indicating that the anisotropy of the Fermi surface and velocity underlies transport anisotropy in the electronic nematic state.
However, the theoretical calculations showed limited quantitative agreement with the experimental data:
the magnitude and in-plane-field anisotropy of the magnetoresistance were underestimated.
Future studies should consider the following possibilities:
(1) The Fermi pockets may be more anisotropic than predicted by density-functional calculations.
(2) The orbital contents on the Fermi surface need to be dealt with explicitly in conductivity calculation. 
(3) Magnetic-field effects on the electronic structure and antiferromagnetic scattering need to be clarified.

The present work demonstrated the effectiveness of investigating electronic anisotropy through interlayer resistivity measurements on Q2D electron systems under magnetic fields.
This new methodology is complementary to surface-sensitive probes such as angle-resolved photoemission spectroscopy and scanning tunneling microscopy.
The present study was mostly performed in the antiferromagnetic state where the nematic order coexists with the magnetic one.
It is highly desirable to apply the present methodology to FeSe in future to study a purely nematic state.

\begin{acknowledgments}
This work was supported in Japan by Japan Society for the Promotion of Science KAKENHI (No. 19H01842, 19H05825, 19H05819, 22H04485, 22K03537).
This work was supported in China by the Youth Innovation Promotion Association of the Chinese Academy of Sciences (No. 2015187).
\end{acknowledgments}

\appendix
\section{Materials and measurements}
CaFeAsF single crystals were prepared in Shanghai by a CaAs self-flux method \cite{Ma15SST}.
Samples with the longest dimension along the [110]$_t$ axis were cut from the grown crystals using a precision wire saw.
Electrical contacts were spot-welded on the crystals and then reinforced with conducting silver paint.
The in-plane resistivity was measured along the longest direction of the samples.
For interlayer resistivity measurements, a current contact and a voltage contact were attached to each (001) plane [see schematic in Fig. 2(b)].
For elastoresistance measurements, a sample was glued on a piezostack, and an electrical current and a strain were applied along the [110]$_t$ axis \cite{Chu12Science, Terashima20PRB}.

The differential thermal contraction method reported in \cite{He17NatCommun, He18PRB} was used to detwin the samples:
A sample was fixed on a PEEK or quartz substrate with stycast 2850FT epoxy encapsulant applied at both [110]$_t$ ends.
The thermal contraction of PEEK (quartz) from 300 to 4 K is -1.1\% (0.03\%) \cite{He18PRB, Nakajima21PRM}.
For comparison, as we do not have values for CaFeAsF, the thermal contraction of SrFeAsF is -0.69 and 0.10\% along the $a_o$ and $b_o$ axes, respectively (Fig. 4 of \cite{Tegel08EPL}).
Because of the differing thermal contractions, when a sample is fixed on a PEEK (quartz) substrate, the fixed direction becomes the shorter $b_o$ (longer $a_o$) axis through most of the sample volume below the structural transition temperature $T_s$ [Fig. 1(a)].
Comparing the resistivity and elastoresistance data of the detwinned samples, we estimated for either substrate that a strain of the order of 10$^{-3}$ was induced at temperatures near $T_s$. 
This strain is comparable to the orthorhombic distortion $\delta = (a_o - b_o)/(a_o + b_o) = 3.4 \times 10^{-3}$ \cite{Xiao09PRB_CaFeAsF, Terashima18PRX}.
The degree of detwinning is difficult to determine for every sample, but the sample fixed on the PEEK substrate for interlayer resistivity measurements reported in Figs. 2 and 3 was almost fully detwinned as confirmed by the $\phi$ dependence of the resistivity at $\theta$ = 90$^{\circ}$ [Fig. 3(f); see above for explanation].
This sample was identical to sample \#1012 in \cite{Terashima22npjQM}.

Magnetoresistance was measured on interlayer-resistivity samples using a
17-T superconducting magnet and a $^4$He variable temperature insert.
The samples were mounted on a two-axis rotator to enable control of both the polar $\theta$ and azimuthal $\phi$ angles of the magnetic field.
$\theta$ and $\phi$ were measured from the $c$ and $a_o$ axes, respectively [Fig. 3(b)].

\section{Theoretical calculations}

The theoretical conductivity $\sigma_{ij}$ ($i=x,~y,~z$) was determined using Chambers' formula \cite{Chambers52ProcPhysSocA}:
\begin{equation}
\sigma_{ij}={e^2 \over 4\pi^3}\int d^3{\bm k}~ \left(-{df({\bm k})\over d\varepsilon_{\bm k}}\right) \varv_i ({\bm k}) \overline{\varv_j ({\bm k})} \tau,
~~~~~
\overline{\varv_j ({\bm k})}={1\over\tau}\int_{-\infty}^0 \varv_j \big({\bm k}(t)\big) \e^{t/\tau} dt,
\end{equation}
where $e$ is the elementary charge, $f$ is the Fermi distribution function with the temperature set to $T$ = 4 K, $\varv_i$ denotes the $i$ component of the quasiparticle velocity $\bm v$, and $\tau$ is the constant relaxation time.
The $x$, $y$, and $z$ axes were set parallel to the $a_o$, $b_o$, and $c$ axes, respectively.
Under a magnetic field $\bm B$, the wavevector ${\bm k}(t)$ at time $t$ is given by the equation of motion:
\begin{equation}
\hbar {d{\bm k}\over dt}=-e {\bm v}({\bm k}) \times {\bm B},
~~~~~{\bm v}({\bm k})={1\over\hbar}{d\varepsilon_{\bm k}\over d{\bm k}},
\end{equation}
which describes the cyclotron motion of quasiparticles.

The band structure $\varepsilon_{\bm k}$ was obtained by Wannier fitting of the FLAPW band structure reported in \cite{Terashima18PRX}.
The Fermi surface is very small and consists of a hole cylinder around the $\Gamma$ point and two electron cylinders [Fig. 1(a)].
The calculated hole (electron) band was shifted slightly by $-24.3$meV ($+8.3$meV) so that the carrier number was consistent with the experimental estimate.
When computing Eqs. (B1) and (B2), we discretized the first Brillouin zone with $1024\times1024\times64$ meshes and summed the contributions from $128\times128\times64$ meshes centered at the hole (electron) surface to obtain the hole (electron) conductivity.
The relaxation time was estimated to be $\tau = 2.06\times 10^{-13}$ s from the Shubnikov--de Haas oscillations of the electrons \cite{Terashima22npjQM}.

The resistivity $\rho_{ij}$ was obtained by tensor inversion of $\sigma_{ij}$.
We found that a simple scalar inversion $\rho_{zz} = 1/\sigma_{zz}$ holds to within the numerical accuracy.
This relation is reasonable, as the Fermi surface consists of cylinders elongated along the $c$ axis.
Accordingly, we defined the electron and hole resistivities along the $z$ axis by $\rho_{zz}^{e(h)}=1/\sigma_{zz}^{e(h)}$.

%

\begin{figure*}
\includegraphics[width=16cm]{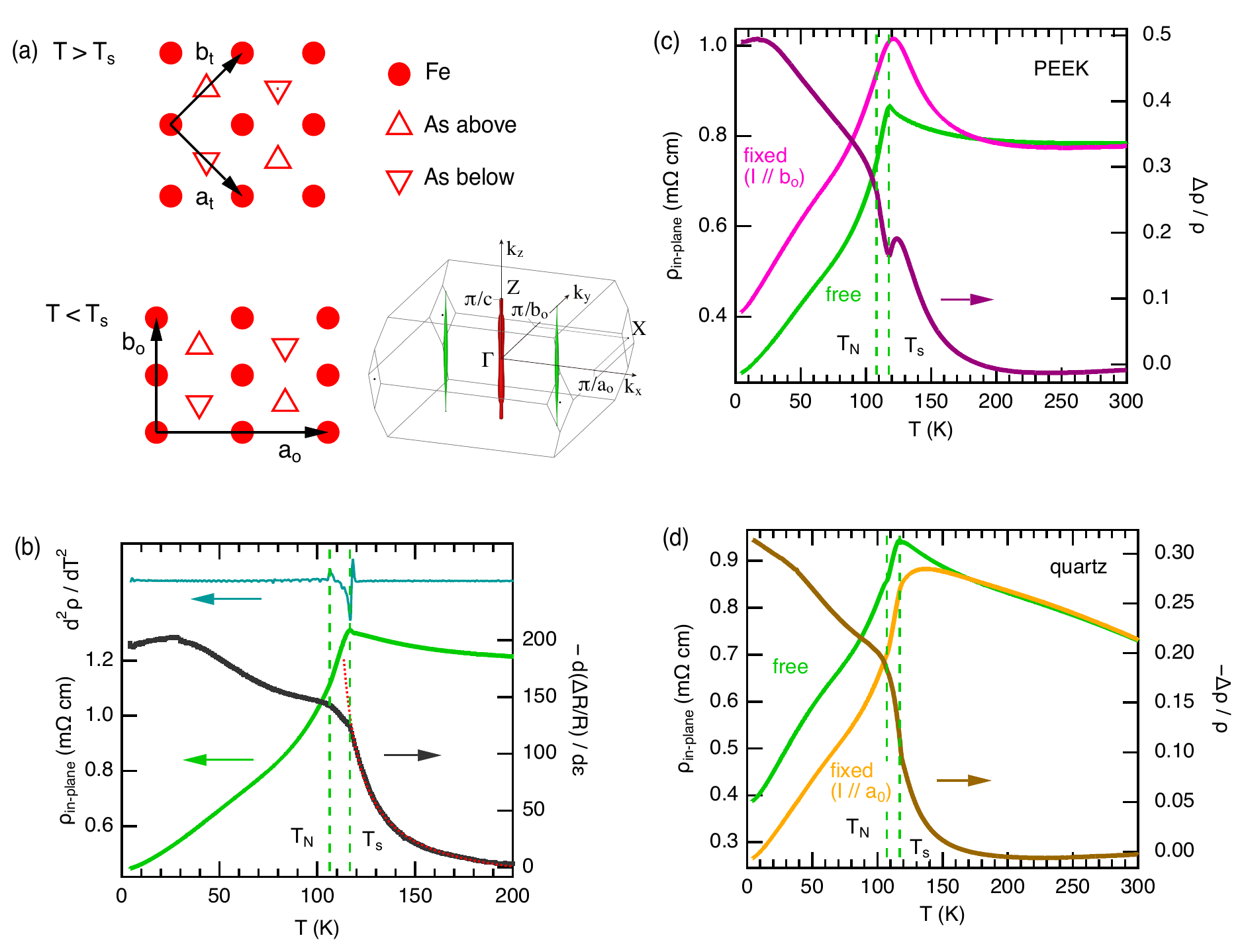}
\caption{\label{Fig1}
In-plane resistivity anisotropy in CaFeAsF:
(a) Schematics of an FeAs layer above and below $T_s$ (the orthorhombic distortion is exaggerated).
Notice that a tetragonal [110]$_t$ axis becomes an orthorhombic $a_o$ or $b_o$ axis below $T_s$.
The lower-right part of this panel shows the Fermi surface in the antiferromagnetic Brillouin zone \cite{Terashima18PRX}.
(b) In-plane resistivity along [110]$_t$ (green), second temperature derivative (light blue), and elastoresistance (black).
The electrical current was applied along [110]$_t$.  
The strain for the elastoresistance measurements was applied along the same direction.
The dotted red line is the Curie--Weiss fit to the elastoresistance. 
The vertical dashed lines represent $T_s$ and $T_N$ determined from the second derivative.
(c) and (d) In-plane resistivity along [110]$_t$ before (green) and after fixing a sample to a PEEK [(c), pink] or quartz substrate [(d), amber].
When fixed, the [110]$_t$ axis becomes mostly the orthorhombic $b_o$ (c) or $a_o$ axis (d) below $T_s$ because the thermal contractions of the sample and substrate differ.
The normalized resistivity difference $\Delta \rho/\rho = (\rho_{\mathrm{fixed}} - \rho_{\mathrm{free}})/ \rho_{\mathrm{free}}$ is also shown (purple or brown).
$T_s$ and $T_N$ were determined before fixing the samples.
}
\end{figure*}

\begin{figure*}
\includegraphics[width=16cm]{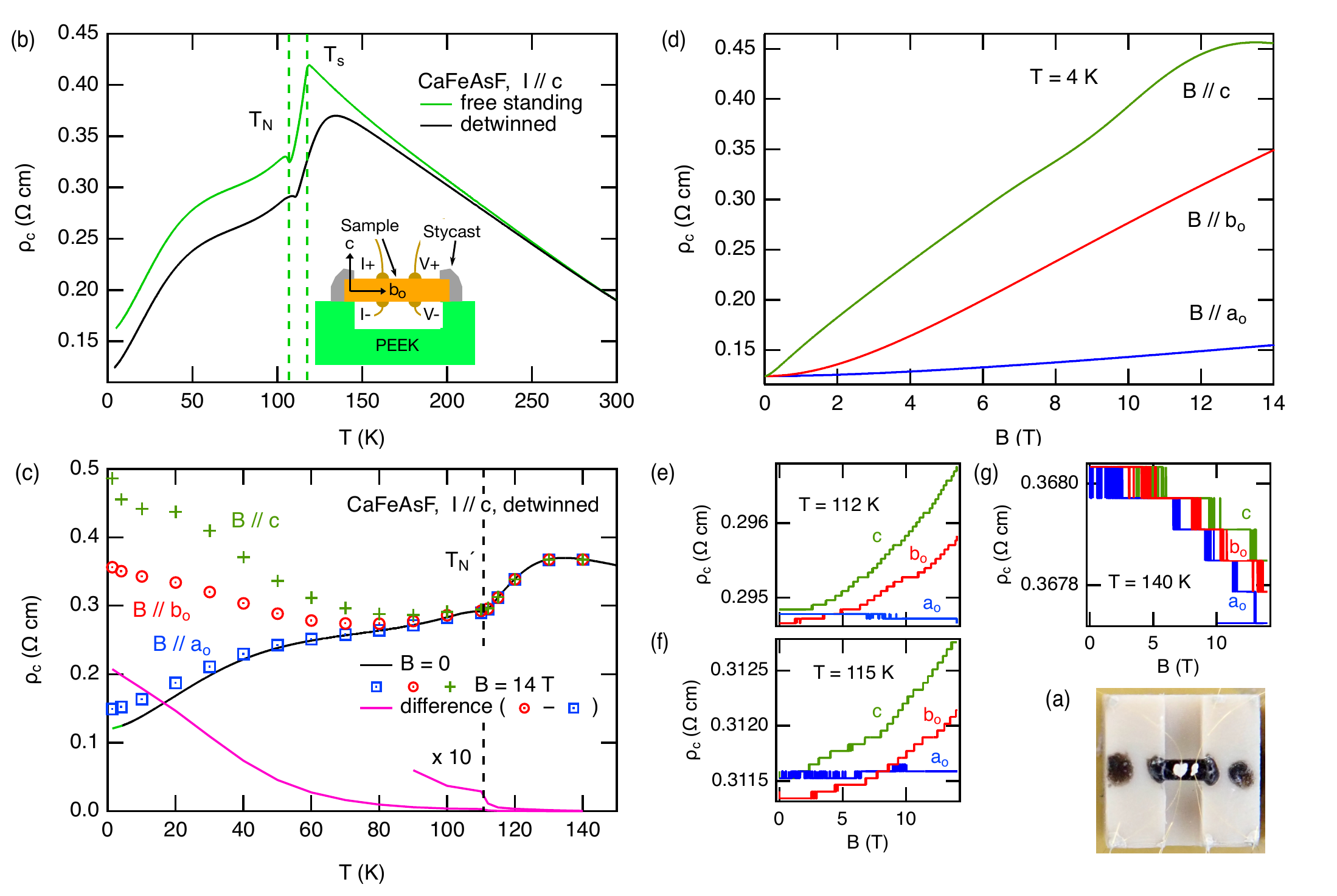}
\caption{\label{Fig2}
Interlayer resistivity in CaFeAsF:
(a) Image of the interlayer resistivity setup.
(b) Interlayer resistivity before (green) and after (black) fixing a sample to a PEEK substrate. $T_s$ and $T_N$ were determined from the second temperature derivative before fixing the sample.
The inset shows a schematic of the setup.
(c) Interlayer resistivities under a magnetic field of 14 T applied along the $a_o$, $b_o$, and $c$ axes (squares, circles, and crosses, respectively), compared with the zero-field resistivity (black line).
The pink curve shows the difference between the $B \parallel b_o$ and $B \parallel a_o$ resistivities.
A ten-times enlarged curve (x 10) is also show in a temperature range between 90 and 140 K.
The vertical dashed line indicates $T_N^{\prime}$ determined after fixing the sample.
(d), (e), (f), and (g) Magnetic-field dependence of the interlayer resistivity under $B \parallel a_o$ (blue), $b_o$ (red), and $c$ (green) measured at $T$ = 4, 112, 115, and 140 K, respectively.
}
\end{figure*}

\begin{figure*}
\includegraphics[width=16cm]{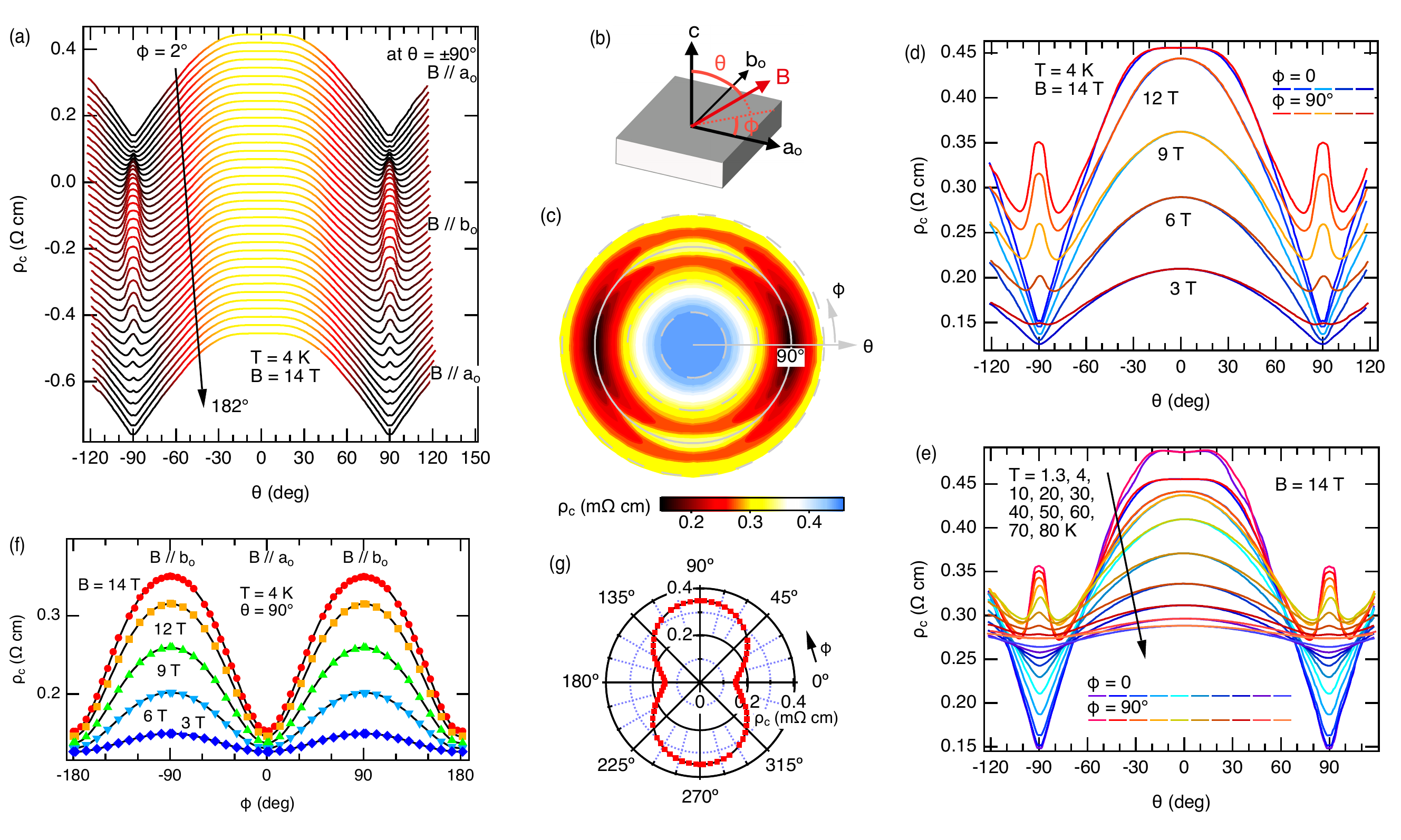}
\caption{\label{Fig3}
Dependence of interlayer resistivity on magnetic-field direction:
(a) Dependence of interlayer resistivity on the polar angle $\theta$ measured under a constant field of 14 T at $T$ = 4 K.
The azimuthal angles ($\phi$) range from 2 (top) to 182$^{\circ}$ (bottom) in steps of 5$^{\circ}$;
the curves are offset for clarity.
(b) Definitions of the field angles $\theta$ and $\phi$.
(c) Azimuthal equidistant projection of the data in (a).
The gray circles represent $\theta$ = 30, 60, 90 (solid) and 120$^{\circ}$.
(d) Dependence of interlayer resistivity on $\theta$ under different magnetic fields (indicated) at $T$ = 4 K and $\phi$ = 0 and 90$^{\circ}$.
(e) Dependence of interlayer resistivity on $\theta$ at different temperatures (indicated) with $B$ = 14 T and $\phi$ = 0 and 90$^{\circ}$.
(f) Dependence of interlayer resistivity on $\phi$ under different magnetic fields (indicated) at $\theta$ = 90$^{\circ}$ and $T$ = 4 K.
The resistivity at $\theta$ = -90$^{\circ}$ is considered to be equal that at $\theta$ = 90$^{\circ}$ and $\phi -180^{\circ}$.
The solid lines are fitted to Eq. 1.
(g) Polar plot of the $B$ = 14 T data in (f) and the corresponding data fit (black line).
}
\end{figure*}

\begin{figure*}
\includegraphics[width=16cm]{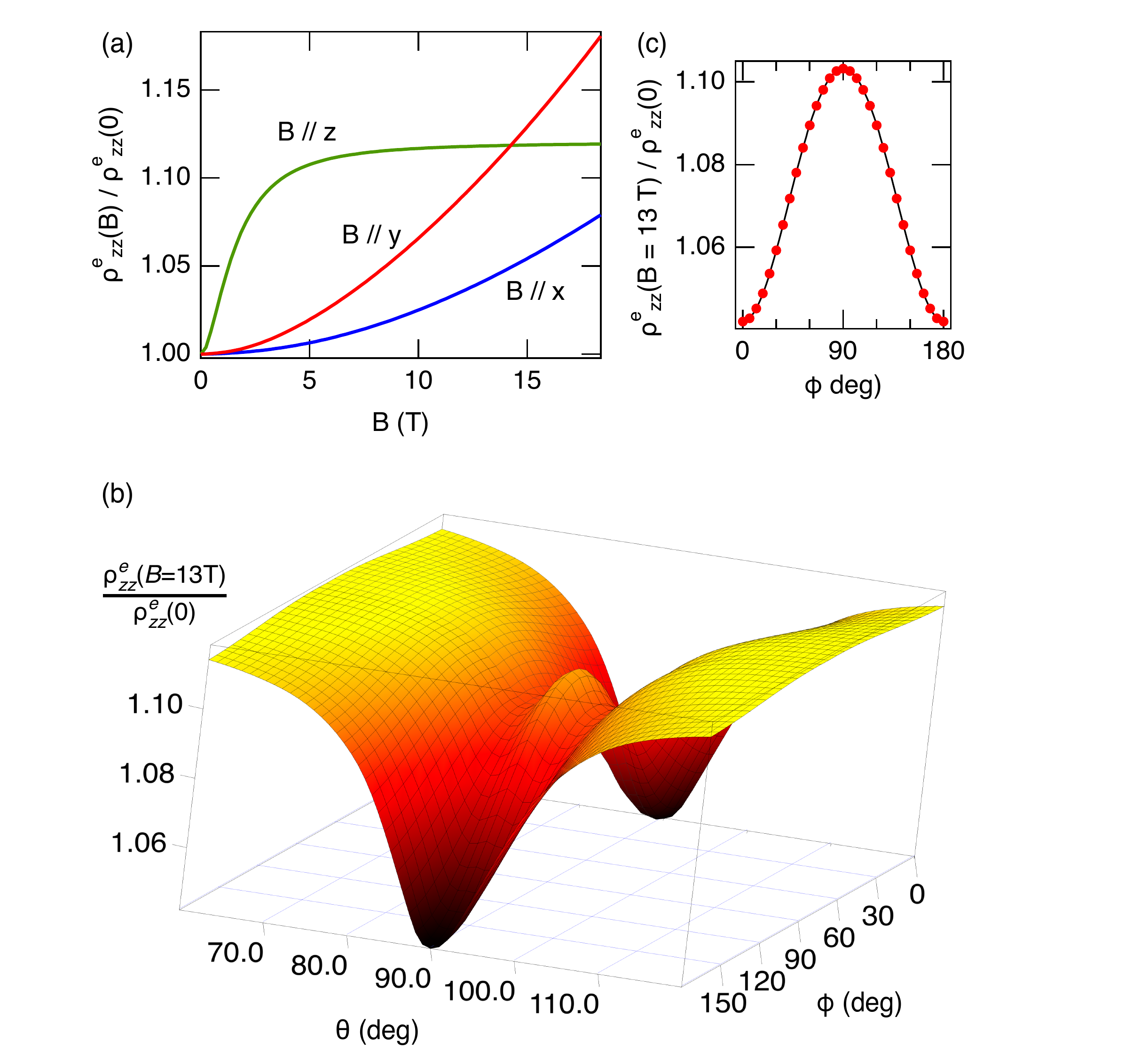}
\caption{\label{Fig4}
Calculated electron interlayer resistivities:
(a) Dependence of electron interlayer resistivity on magnetic fields $B \parallel x$, $y$, and $z$.
(b) Dependence of electron interlayer resistivity on $\theta$ and $\phi$ near $\theta$ = 90$^{\circ}$ for $B$ = 13~T.
(c) Dependence of electron interlayer resistivity on $\phi$ at $\theta$ = 90$^{\circ}$ for $B$ = 13~T.
The solid line is fitted to Eq. 1.
}
\end{figure*}

\begin{figure*}
\includegraphics[width=16cm]{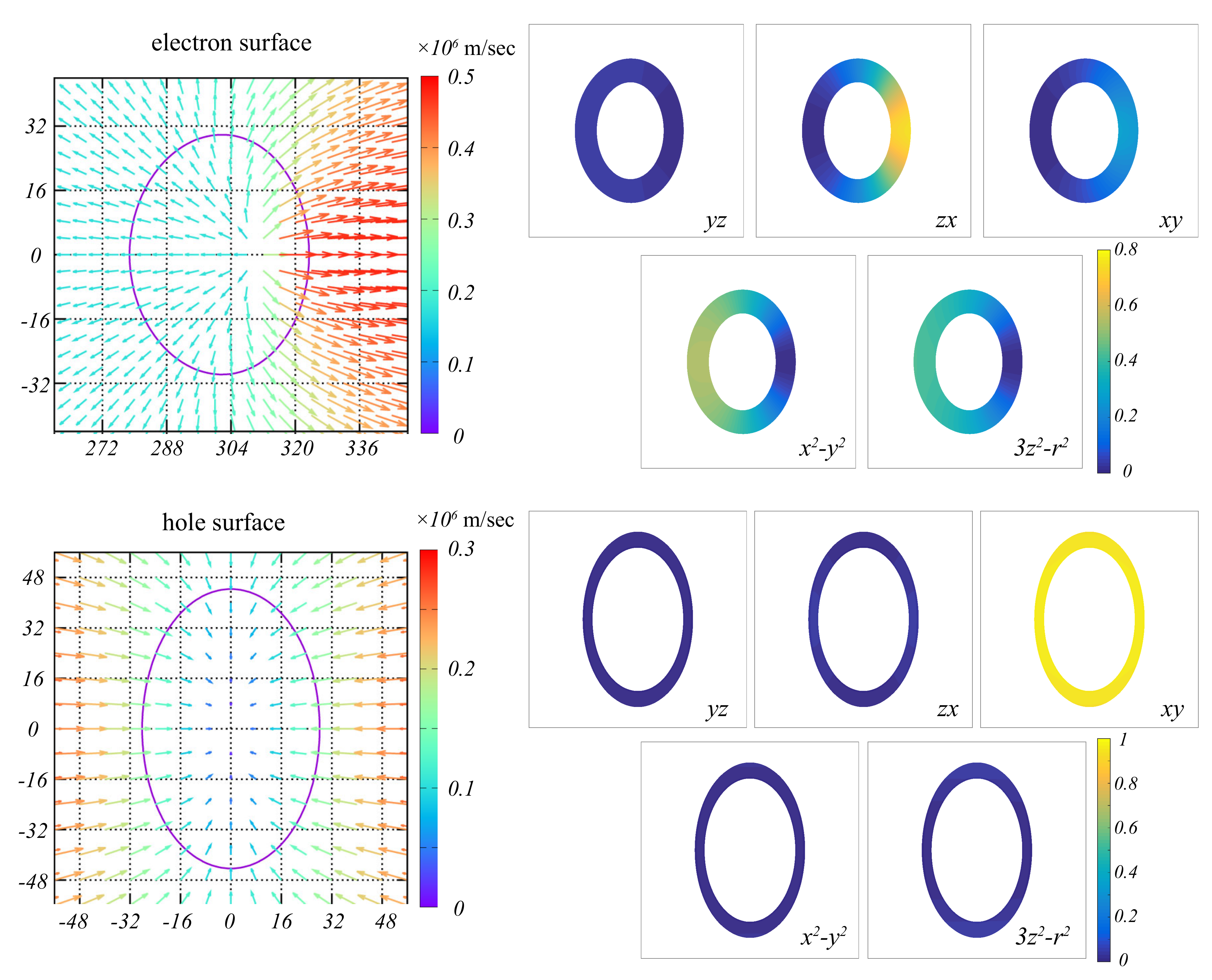}
\caption{\label{Fig5}
Velocity map and orbital contents of the electron and hole Fermi surfaces.
In all subfigures, the horizontal direction is parallel to the $k_x$ direction.
The subfigures on the left show the velocity distributions $(1/\hbar)\nabla_k \varepsilon(k)$ around the cross-sections at $k_z$ = 0 for the electron (upper) and hole (lower) surfaces.
The numbers attached to the horizontal and vertical edges refer to the $1024\times1024$ meshes of the Brillouin zone.
The other subfigures show the orbital contents along the cross-section.
}
\end{figure*}

\end{document}